# LA MUJER A TRAVÉS DE LOS PERSONAJES FEMENINOS EN EL CINE DE TEMÁTICA FINANCIERA

Por Inés Martín de Santos

**Resumen / Abstract**

Análisis y valoración de los personajes femeninos en los filmes de temática económico-financiera aparecidos en una treintena de películas relevantes, desde los inicios del cine, seleccionadas de varios repertorios. Estudio descriptivo, comparativo e imparcial. Se aplican técnicas cuantitativas para medir la influencia de la mujer en cuanto a protagonismo, valores positivos, valores negativos, acatamiento o incumplimiento del test de Bechdel, y se evalúan resultados. Se parte de dos publicaciones previas vinculantes: Marzal et al. *La crisis de lo real. Representaciones de la crisis financiera de 2008 en el audiovisual contemporáneo* (2018) y Martín-de-Santos, *La administración financiera a través del cine. Estudio analítico y crítico de las películas más relevantes para la enseñanza universitaria* (20219. Se contrastan las versiones cinematográficas de carácter realista con los hechos reales en la actualidad.

> “Que les femmes considerées sélon les principes de la saine Philosophie sont autant capables que les hommes de toutes sortes de connoissances [sic].
> “Il est aisé de remarquer, que la difference des sexes ne regarde que le Corps: n'y ayant proprement que cette partie qui serve à la production des hommes; & l'Esprit ne faisant qu'y préter son consentement, & le faisant en tous de la même maniere, on peut conclure qu'il n'a point de sexe”[1].
>
> (François Poullain de la Barre. *De l'égalité des deux sexes, discours physique et moral où l'on voit l'importance de se défaire des préjugez 1673*, pp. 109-110).

## Introducción

[1] Las mujeres, consideradas según los principios de la sana Filosofía, son tan capaces como hombres de todo tipo de conocimientos. Es fácil notar que la diferencia entre los sexos concierne únicamente al Cuerpo: estrictamente hablando sólo existe la parte que sirve para la reproducción de las personas; el Espíritu sólo es para dar su consentimiento, y haciéndolo de la misma manera, podemos concluir que no tiene sexo.

La participación de las mujeres en la ciencia y en la creación artística se ha producido en cualquier período de la historia de la humanidad desde las primeras culturas. En la Edad Moderna, el hecho de no haberlas prestado tanta atención como hubiera sido y es deseable, más que a una actitud premeditada por ocultar tal evidencia (salvo algunos casos), se ha debido, al menos en parte, al escaso estudio y consiguiente desconocimiento de esta realidad.

En España sin ir más lejos, por ejemplo, en el campo de la literatura, es asombrosa la producción femenina que no aparece en los manuales. No hay más que consultar los principales repertorios de Manuel Serrano y Sanz (1903-5), Aurora Díaz-Plaja (1945), Eduardo Martín de la Cámara y Diego Ignacio Parada y Barreto (1920), Carmen Simón Palmer (1991) o Lidia Falcón (1992) para darnos cuenta de la enorme, y de momento inabarcable, contribución de las mujeres a nuestra literatura.

Sin embargo, en el cine, la mujer no ha tenido una influencia tan determinante hasta hace unas décadas, entre otros motivos por haber prevalecido su condición de actriz sobre la de autora en sus diversas facetas (productora, guionista, …) pese a la trascendencia de mujeres excepcionales como Alice Guy. De hecho, todavía hoy, en nuestro país, aproximadamente el 68% de los guionistas son varones frente al 32% de mujeres. Hay que suponer que esta brecha se irá reduciendo con el tiempo. En cuanto a la dirección de películas, la posibilidad de que las mujeres produzcan y dirijan trabajos cinematográficos, se ha ampliado a medida que estos se están desarrollando en plataformas como Netflix, Amazon, Hulu y Apple, pero hasta hace poco, según un estudio reciente (McIntyre, 2021), se ha visto que, en las 250 películas mejor valoradas por la crítica, tan solo aparecen 4 directoras.

¿De qué manera se ha idealizado la mujer en el cine? Tenemos estudios de tipo generalista (Haskell, 1974; Arranz, 2010, …) o enfocados desde diversas perspectivas, las más recurrentes son: la mujer en el cine mudo, mujeres tratadas por determinados directores, la mujer en el mundo hindú, la mujer en el islam, la mujer en el cine de tal o cual país (Polonia, etc.), mujer y violencia, la mujer en el ámbito laboral, mujer en la política, mujer y erotismo, la mujer en tiempos de guerra, vampiresas, religiosas, amas de casa, … El propósito de esta investigación es original y consiste en ver concretamente el grado de incidencia de los personajes femeninos en las películas de contenido financiero con el fin de contrastarlos con la influencia de las mujeres en el mundo real.

Desde finales del pasado siglo, se viene observando una incipiente aparición cinematográfica de la mujer en la esfera de los negocios como protagonista tal como consta en algunos títulos como *Working Girl* (1988), *Small Time Crooks* (2000), *Mostly Martha* (2001), *Coco Chanel* (2008), *Sunshine Cleaning* (2008), *Julie & Julia* (2009), *She ++ the documentary* (2013), *Born to* Win (*Nacida para ganar*, 2016), *The bookshop* (2017), *The Ground Beneath My Feet* (2019 y, cómo no, asimismo vemos filmaciones españolas dignas de mención como *7 mesas de billar francés* (2007) de Gracia Querejeta.

Curiosamente la relevancia de la mujer en la economía, desde la perspectiva cinematográfica, también ha proliferado en algunos países en vías de desarrollo y ha presentado algunos productos de calidad tales son los casos de *Mother India* (India, 1957) de Mehbood Khan, sobre la usura a la mujer prestataria; *Adam* (Marruecos, 2019) de Maryam Touzani, sobre la mujer emprendedora; o la muy graciosa *Eu tu eles* (Brasil, 2000) de Andrucha Waddington, sobre la mujer campesina desprovista de prejuicios.

*Grosso modo*, como más adelante se demostrará, la escasa atención a las mujeres en las películas de asuntos financieros no hace justicia a su importancia en la vida real, al menos si se tiene en cuenta que la incorporación de la mujer al trabajo, y sobre todo al trabajo cualificado, está cambiando progresivamente las relaciones personales y está incidiendo notablemente en los modelos económicos.

Esta clase de cine, incluso la tendente a la ciencia-ficción, no ha sabido desprenderse adecuadamente de determinadas costumbres asentadas en la tradición y de ciertos condicionamientos sociales, por ello los personajes femeninos desempeñan mayoritariamente papeles secundarios y esto es así, entre otros motivos, porque de algún modo abundan las escenas retrospectivas anquilosadas en el pasado, así como también se reflejan realidades de viejos tiempos como, por ejemplo, las manidas imágenes sobre la bolsa, donde a menudo aparecen frecuentemente los tradicionales corros de compra y venta de títulos a viva voz en el parquet.

En el entorno financiero real, el protagonismo femenino es evidente. Observamos casos harto significativos como la concesión de microcréditos por parte del banquero Muhammad Yunus exclusivamente a las mujeres, probablemente porque estas ofrecen un mayor grado de fiabilidad y menor riesgo que los hombres para la devolución de los préstamos. Por contra, en los rodajes de Hollywood, las actrices han sufrido una discriminación salarial respecto a los actores (Izquierdo Sánchez; Navarro Paniagua, 2017).

Si se tratara de películas sobre finanzas, la reciente tendencia a no diferenciar premios para actores y premios para actrices en algunos festivales de cine daría unos resultados poco equitativos, puesto que rara vez estas filmaciones están protagonizadas por mujeres. Algo similar sucede con las actuales normas de descripción bibliográfica para averiguar la producción científica de las mujeres al desaparecer el nombre de pila completo en el campo de autoría, lo que ocasiona escollos para realizar estudios bibliométricos de género.

## Metodología

Se han escogido 30 películas utilizando como referencia las listas de Mateer; O'Roark; Holder (2016), Leet y Houser (2003), Casares Ripoll, Ramos Gorostiza y Santos Redondo (2004), Guivernau Molina (2013), Torres Dulce (2014), Sánchez-Pagés (2019) y mi modesta proposición (Martín-de-Santos, 2021).

El criterio de selección ha sido en primer lugar, figurar repetidas veces en al menos tres de las seis fuentes citadas; en segundo lugar, responder a un cierto grado de calidad artística y, en tercer lugar, atender a su popularidad en España. En este sentido se han descartado trabajos de mérito, pero más discretos como, por ejemplo, *The Pursuit of Happyness* (*En busca de la felicidad*, 2006) o *Rollover* (*Una mujer de negocios*, 1981) que, pese al interés del tema, no fue muy convincente ni tuvo demasiada aceptación tanto por parte de la crítica como del público.

El período de aparición es largo para observar, de este modo si se han producido algunos cambios de sensibilidad en los directores, en los gustos del público y en las costumbres de época. Por ejemplo, es raro no ver escenas de fumadores en las películas del cine negro norteamericano de mediados del pasado siglo, frente a su progresiva disminución en la actualidad.

Se han eludido las grabaciones de carácter documental por no tratarse de productos destinados a la imaginación y, por lo tanto, no esencialmente creativos como *Tulipmanía* (2000), *Enron. Los tipos que estafaron a América* (2005), *Collapse* (2009), *Inside Job* (2010), *The flaw* (2011), …

Al no tocar directamente el ámbito financiero, ha sido perentorio excluir películas en las que la protagonista femenina es el centro de la trama como *Jonhny Guitar* (1954) con una Joan Crawford que recuerda mucho a las mujeres vestidas de hombre en el teatro español durante el siglo XVII, o *The Children's Hour* (*La calumnia,* 1961) cuyo contenido fundamental se centra en la falta de moralidad de una alumna más que en la ruina económica del centro de enseñanza que ello supone, o la irregular *Promising Young Woman* (*Una joven prometedora*, 2020), recomendable para asumir la equidad de género.

Quedan en el tintero algunos productos con los que la gente occidental no suele estar tan familiarizada, pese a su calidad, procedentes de países como, por ejemplo, en Rusia *Una rubia a la vuelta de la esquina*, de Vladímir Bortko (1984) o las chinas (Jia, 2017) *La esposa abandonada* de Hou Yao (1924), *La profesión de la señora Wei* de Zhang Shichuan (1927), *Derechos de la mujer* del mismo director (1936), ... y películas más recientes muy dignas de considerar como *Chocolat* (2000) por limitación de la muestra, y que evidencian el progresivo protagonismo de las mujeres en el emprendimiento económico.

La estructura de la muestra abarca las categorías: 1. mujer protagonista, 2. Mujer de reparto. 3. Atributos positivos, 4. Atributos negativos, 5. Cumple el test de Bechdel.

Los criterios para señalar atributos positivos son cualidades como: bondad, honestidad, generosidad, fidelidad y equivalentes. Criterios para señalar atributos negativos: maldad, deshonestidad, avaricia, infidelidad, delincuencia y similares. En los casos en que en una película aparezcan mujeres con valores positivos y mujeres con valores negativos, se excluye el valor positivo o negativo de menor trascendencia en la trama como ocurre en el nº 19 de la lista presentada más adelante.

Recurro al test de Alison Bechdel para implementar una de las variables inmersas en la muestra con el fin de evaluar la brecha de género.

Para admitir si se cumple o incumple el test de Bechdell se considera que los diálogos entre mujeres que no estén relacionados con asuntos de hombres tendrán una duración razonable y de, al menos, cuatro o cinco minutos. Una conversación entre dos mujeres en una película en la que ese diálogo no influya decisivamente en la trama general o en el desenlace de una acción destacable, no se considera reconocida por el test de Bechdell.

Se trata de un estudio descriptivo, cuantitativo y comparativo. El propósito general de esta investigación es contrastar el tratamiento de los personajes femeninos en el cine con la conducta de las mujeres en la vida real. En este sentido, desde una perspectiva diacrónica, se trata de ver qué películas reflejan fielmente la realidad y cuáles la deturpan.

## Selección de películas

Se presenta una lista de los filmes con breves indicaciones que se estiman relevantes para el objetivo principal del estudio. Las obras recabadas, por orden cronológico, son las siguientes:

1- *Die Finanzen des Großherzogs* / *Las finanzas del Gran Duque* (1924). Extraordinaria comedia del cine mudo repleta de detalles sorprendentes y escenas conmovedoras. Nicolai, el alegre y descuidado Gran Duque de Abacco (isla mediterránea) se encuentra arruinado. Recibe una carta de la Gran Duquesa Olga en la que le comunica su deseo de casarse con él por haber ayudado a los náufragos del Palestrina y de paso, le comunica que remediará su banca rota. Tras diversas argucias de amigos y enemigos del Duque, la trama terminará con un final feliz.

2- *American Madness* / *La locura del dólar* (1932). Llamativo título para una intriga de zancadillas entre los directivos de un banco. Tom Dickson (Walter Huston), honesto director de un banco y romántico hombre de negocios, durante el período de la gran depresión, es citado por la junta directiva para fusionar esta entidad con el New York Trust, a lo que él se niega. Entre tanto, un empleado, acorralado por deudas de juego, es acosado por mafiosos para cometer un robo en el banco y les facilita la sustracción. Las sospechas recaen sobre otro trabajador exconvicto que, por fidelidad a Dickson, no quiere denunciar al verdadero responsable del robo que durante el atraco pilló coqueteando con la mujer de Dickson. Ante este acontecimiento, los clientes sacan sus ahorros del banco, pero lo amigos de Dickson realizan importantes imposiciones que salvan la debacle. Excelente película de Frank Capra con un final ingenuo.

3- *Modern Times* / *Tiempos modernos* (1936). Obra maestra del sétimo arte. Espectacular teatralización de Charles Chaplin plaga de detalles muy interesantes como la fotografía de Lincoln en la celda de Chaplin. Exagerada interpretación del desarrollo industrial (para potenciar los rasgos de humor), los fallos de la justicia y, en definitiva, la deshumanización de las personas. Un obrero de fábrica en cadena enloquece por el frenético ritmo de trabajo, ingresa en un hospital y a la salida es detenido por participar inopinadamente en una manifestación. En la cárcel, ayuda de manera involuntaria a sofocar un motín y ya en libertad reanuda su lucha por la vida junto a una joven huérfana con la que, tras desafortunados avatares, termina juntándose y ambos desaparecen andando por una carretera.

4-*The Grapes of Wrath / Las uvas de la ira* (1940). Otra obra maestra. El escéptico e inconformista John Ford presenta la desposesión de tierras a las familias por parte de los bancos y de las grandes compañías anónimas que obligan a los campesinos a buscar oportunidades en otras tierras durante la gran depresión norteamericana. No falta la policía ruin para añadir mayor emoción a la historia. Imágenes pintorescas del medio rural como los burros empleados para el trabajo y lenguaje poético conmovedor. El protagonismo y enaltecimiento de las mujeres adquiere carácter monográfico, no obstante, en *7 mujeres*, última película de este genial director.

5- *It's a wonderful life* / *¡Qué bello es vivir!* (1946). Segunda gran película de Frank Capra en la que mezcla problemas financieros con relaciones sentimentales y el triunfo final de estas últimas. Comedia amena y algo lacrimógena en la que un banco comercial tiene que hacer frente al oligopolio del magnate de la ciudad que no duda en emplear artimañas para hundirlo ante la resistencia del primero a ser absorbido, con mezcla de elementos fantasiosos como la intervención de un ángel de la guarda.

6- *The Man in the White Suit* / *El hombre del traje blanco* (1951). Película que no ha perdido actualidad en los tiempos de la obsolescencia programada de los productos. El joven Sid Stratton inventa una tela que ni se rompe ni se mancha ante el estupor de los fabricantes, que verán mermada la producción, y de los empleados que verán amenazados sus puestos de trabajo.

7- *Bienvenido mister Marshall* (1953). Parodia del plan norteamericano de ayuda a las economías europeas dañadas por la Segunda Guerra Mundial. Como la gran mayoría de las ayudas no es gratuita, esta fue la antesala y una de las mejores estrategias para la exportación de productos norteamericanos que sirvieron a EEUU para remontar la gran depresión. Los papeles de la maestra del pueblo, la señorita Eloísa (Elvira Quintilla) y Carmen Vargas (Lolita Sevilla) fueron secundarios pese a haber sido esta última la mejor pagada del reparto y pese a aparecer en primer lugar en la presentación por requisito del productor. El alcalde de Villar del Río, don Pablo (Pepe Isbert) y el contratista de cuadros flamencos Manolo (Manolo Morán) interpretaron los papeles principales. Mágica manera de presentar un simpático entorno rural español. Escenas en su momento censuradas como la bandera estadounidense arrastrada por las aguas. Mordaz crítica encubierta que la censura no percibió.

*8- Executive Suite* / *La torre de los ambiciosos* (1954). Rifirrafe, contraprestaciones y sobornos entre los directivos de una empresa fabricante de muebles por la sucesión a la presidencia. La presencia femenina se reduce a la esposa de un directivo, a la amante de otro, a las secretarias de la empresa y a la de la accionista mayoritaria, Julia O. Tredway (Barbara Stanwyck), cuyo peso en el consejo es decisivo para que la elección recaiga en McDonald Walling (William Holden) defensor de la calidad frente a la desmedida especulación. Buena película que mantiene en todo momento la tensión de la trama y reclama la atención del público y a ello ayuda, en este caso, la ausencia de banda musical.

9- *Ziemia obiecana* / *La tierra de la gran promesa* (1975). Frialdad del capitalista imperturbable ante las desgracias humanas: que se manche una tela de sangre, por ejemplo, es peor que un hombre pierda el brazo en una máquina. Frases como "yo no me puedo permitir el lujo de sufrir con cada perro de la calle" dan idea de la indolencia que caracteriza fundamentalmente este filme. Crítica feroz al capitalismo abusivo. Tres jóvenes, uno católico, otro luterano y otro judío, montan una fábrica textil en Polonia durante la revolución industrial. Los devaneos amorosos del católico con la mujer de un judío provocan la venganza de este último quemando la fábrica. Tras unos años, por medio de un matrimonio de conveniencia con el católico, la fábrica vuelve a renacer, pero se enfrenta a la huelga de los trabajadores que es reprimida a tiros. A pesar de su larga duración, el vestuario y la espectacularidad de los escenarios convierten toda la predominante majestuosidad en una película de culto. La mujer, sometida a constantes abusos, en especial la mujer obrera, aparece como una pieza más de todo este entramado; en este sentido es una película de denuncia.

10- *Trading Places* / *Entre pillos anda el juego* (1983). Divertida comedia, pese al humor siempre algo forzado de Eddie Murphy, en la que dos magnates hacen una apuesta para ver la reacción de dos personas a las que cambian artificialmente el destino. Convierten a un ejecutivo en mendigo y a un miserable en ricachón con toda clase de comodidades. La broma les cuesta cara porque estos individuos, que casualmente cruzan sus vidas, unen sus esfuerzos finalmente para hundir las acciones de sus poderosos jefes en la bolsa.

11- *Wall Street* (1987). El título alude a la calle donde se ubica la bolsa de Nueva York, el mayor mercado de transacciones dinerarias en del mundo y meca de los inversionistas en renta variable. Con esta película se crea uno de los *locus amoenus* más recurrentes de los contenidos sobre finanzas. El joven bróker Bud Fox (Charlie Sheen) logra entrar en el equipo del gran hombre de negocios Gordon Gekko (Michael Douglas); tras descubrir las artimañas nada ortodoxas de Gekko en la adquisición de una compañía de aviones (mediante una opa) donde el padre de Bud es sindicalista, consigue bajar el precio de las acciones obligando a Gekko a deshacerse de los títulos. Técnicamente bien cuidada, hasta la voz de Frank Sinatra en la presentación es muy acertada. Va a marcar los componentes básicos de la estela de filmes siguientes: Escenas rápidas, lenguaje provocativo, ambición desmesurada, información privilegiada, falta de escrúpulos, lujo, sexo, … Aquí el papel de Darien Taylor (Dary Hannah) la actriz más destacada y oportunista, ocupa un plano secundario.
En el año 2010 se rodó una continuación: *Wall Street 2: Money Never Sleeps* (*Wall Street 2. El dinero nunca duerme*) que no tuvo un eco especial y fue recibida con frialdad por la crítica.

12- *Working Girl* / *Armas de mujer* (1988). La adaptación del título inglés al español no se ajusta fielmente al contenido de la obra, tampoco se debe confundir con la serie televisiva que emplea el mismo marbete (algo que no se debe hacer por ir contra los derechos de autor) protagonizada por Sandra Bullock dos años después. Aunque el hilo conductor sea inicialmente la capacidad profesional de Tess McGill, en realidad se trata de un trasunto híbrido entre negocios y erotismo perfectamente resumido en la frase de Tess "tengo una mente para las finanzas y un cuerpo para el pecado" y con unas sugestivas escenas de lencería femenina. Buen guión bien trabado con abundantes enredos y expresiones afortunadas fuera de contexto que dan pie a la ironía ("poder para el pueblo"). Los protagonistas son Tess McGill (Melanie Griffith) y Jack Trainer (Harrison Ford) aunque como complemento necesario figure asimismo Katharine Parker (Sigourney Weaver). Aunque Tess McGill triunfe en su puesto de trabajo, no deja de ser una empleada al servicio del dueño del dinero.

13- *Other People's Money / Con el dinero de los demás* (1991). En la línea anterior, el contenido, de manera muy sucinta, gira en torno a la adquisición de pequeñas empresas cotizantes en bolsa por el poderoso Larry Garfield (Danny DeVito), en este caso planea una opa a la compañía de Andrew Jogerson (Gregory Peck) mediante la intervención de la abogada Kate Sullivan (Penelope Ann Miller). La obra tiene aires de comedieta. Algunas incongruencias: escenarios demasiado austeros para la importancia de los asuntos tratados. Empresa familiar muy pequeña para cotizar en bolsa. El pizpireto Danny DeVito, con cierto parecido a los drásticos movimientos de James Cagney, no convence ni como hombre de negocios ni como acosador sexual; tampoco Gregory Peck parece un octogenario (edad real 75 años sin maquillar cuando se filmaron las escenas).

14- T*he Hudsucker Proxy / El gran salto* (1994). Película imaginativa en la que se entremezcla una excéntrica realidad con la ficción. Merecedora de mayor atención de la que ha tenido, con buenas dosis de refinada ironía (a veces difícil de percibir), con un experimentado Tim Robbin y coronada por Paul Newman. Inopinadamente fue un fracaso comercial. El argumento versa sobre el neófito economista Norville Barnes (Tim Robbin) el cual entra a trabajar de empaquetador en una empresa donde llega a ser presidente mediante la argucia de uno de los miembros de la junta directiva (Sidney J. Mussburger / Paul Newman) para sustituir al presidente de la compañía que recientemente se había suicidado y con el fin de mantenerlo como perro de paja. La periodista Amy Archer (Jennifer Jason Leigh) es contratada para escribir la historia de Norville, acaba siendo su secretaria y posteriormente su amiga. La pretensión de defenestrar a Norville le sale caro a Mussburger que acaba siendo recluido en un siquiátrico. Velada crítica al mundo de los negocios como la mayoría de estos filmes. Es un trabajo de bastante mayor calidad que las de su estilo.

15- *Rogue Trader / El gran farol* (1999). Basada en hechos reales. El banco Barings encarga al joven Nick Leeson (Ewan McGregor) resolver un problema en su sede de Yakarta, donde se casa con Lisa (Anna Friel). Desde allí le trasladan a Singapur destinado al mercado de futuros. Por su cuenta realiza inversiones de riesgo no autorizadas con cargo a los fondos del banco generando una deuda enorme a esta entidad. Soportó una condena de cuatro años de cárcel. Como suele ser corriente en las películas de esta temática: prisas, llamadas telefónicas, gente trajeada, lujo, muchas discusiones y, en este caso, acciones supuestas suplidas por voces en off. Producto discreto.

16- *Boiler Room / El informador* (2000). Seth (Giovanni Ribisi), hijo de un juez, pasa de organizar timbas ilegales en su casa a trabajar en JT Marlin (chiringuito financiero) con el fin de aprender los entresijos de la bolsa y conseguir al menos 40 cuentas para poder instalarse como corredor independiente. Película de ritmo trepidante con una recalcitrante música hip-hop para enervar al espectador más sereno. Sigue la línea de *Glengarry Glen Ross* (*Éxito a cualquier precio*, 1992) y de casi todas las películas presentadas en esta lista. En el fondo denuncia la agresividad y malas prácticas financieras, pero ¿qué podemos esperar hoy día si un movimiento del jugador de fútbol Cristiano Ronaldo rechazando dos botellas de Coca-Cola por agua en una entrevista es capaz de hacer perder en la bolsa norteamericana cuatro mil millones de dólares?[2] Película discreta. Salvando la aparición de una secretaria, personajes femeninos prácticamente inexistentes.

---

[2] BBC News Mundo, 16/06/2021) https://www.bbc.com/mundo/deportes-57484146 [consulta 18/06/2021]

17- *A beautiful mind* / *Una mente maravillosa* (2001). Trata la vida del premio nobel de Economía John Forbes Nash (Russell Crowe), gran innovador de la Teoría de Juegos, pero a través de la novela de Sylvia Nasar, lo que ha supuesto algunas malinterpretaciones científicas. La famosa escena de las 5 chicas en el bar, por ejemplo, para la posible elección de ellas por los 4 amigos, tergiversa la teoría del equilibrio de Cournot o de Nash, o *Cast Away* cuya relación con la Economía es algo forzada por el mero hecho de que el actor principal sea un analista de sistemas. Tres son los ingredientes principales de la película: ciencia, amor y espionaje. Se pueden disculpar algunas frases prepotentes de los hombres porque la película es ficción y no tiene por qué responder estrictamente a la realidad, pero lo sorprendente es que el protagonista dé clase en un aula donde sólo aparezca una alumna que se convertirá más tarde en su esposa; algo similar sucede con el porcentaje de hombres y mujeres en cuanto al número de extras seleccionados para hacer bulto. Los méritos han sido sobrevalorados.

18- *Die Fälscher* / *Los falsificadores* (2007). Derroche de lujo escénico y un dialogo dinámico no exento de algunos sarcasmos ("—¿Por qué no está Dios en Auschwitz? —Porque no pasó la selección"). A vueltas con los asuntos bélicos, muestra tremendas y terribles imágenes realistas no relativas exclusivamente a los malos tratos en los campos de concentración. Justamente premiada con el óscar a la mejor película de habla no extranjera en 2008. Tras la segunda guerra mundial, en el casino de Montecarlo, un judío rememora la historia de quienes se vieron forzados a ayudar a los nazis a falsificar la libra británica y el dólar estadounidense para hundir la economía de sus correspondientes países. Se corrobora el principio elemental de que el dinero es, si no la fundamental, una de las mejores armas para ganar una guerra.
Salvo escenas familiares de oficiales alemanes y compañía de mujeres en el casino, apenas hay intervención femenina; hasta en la fiesta de carnaval solamente intervienen hombres. La escena final bailando el tango a la orilla del mar no parece muy acorde con el resto.

*19- The Company men* (2010). En la tónica de las películas de bajo presupuesto con excesiva teatralidad; montaje en ocasiones deficiente; el guion repite un argumento manido y un lenguaje similar: el dueño de una empresa naviera incorporada una corporación internacional con una esposa despilfarradora y una amante que es la directora de recursos humanos de la corporación (escenas pacatas e irrisorias). Se confirma la crisis económica del 2008 y la solución según el nuevo CEO es la reducción de plantilla para abaratar costes. Tras el despido, se producen cambios en algunos de sus empleados: desde la adaptación a otros trabajos de menor categoría hasta el suicidio. Los personajes femeninos tienen facetas tanto positivas (una ama de casa se pone a trabajar) como negativas (colaboradoras con la infidelidad del marido).

20- *The Last Days of Lehman Brothers* / *Los últimos días de Lehman Brothers* (2010). Aprovechando la fama de la noticia sobre la quiebra del banco de inversión Lehman Brothers, la BBC produjo esta recreación algo sensacionalista híbrida entre la ficción, la documentación y la reflexión mediante una voz en off que determina su carácter predominantemente narrativo con juicios en ocasiones no muy afortunados y apropiados para el debate en las aulas. La falta de regulación de los mercados y la implicación en las hipotecas *subprime* fueron los motivos principales de la caída de este emporio financiero. La premura en las explicaciones a veces dificulta la comprensión cuyas dudas, en este caso, me parece que se pueden resolver mejor leyendo libros tan clarificadores como el de Matt Taibbi*, Cleptopía, fabricantes de burbujas y vampiros financieros en la era de la estafa*. No obstante, en el minuto 17,40 aparece una explicación de tipo divulgativo ilustrativa y aceptable. Merece la pena ver, como en tantas otras ocasiones, al excelente actor secundario James Cromwell. Escasa participación de mujeres.

21- *Margin Call* / *El precio de la codicia* (2011). Reitero mi valoración en un estudio anterior: "En teoría el título alude a la llamada del bróker para la reposición del depósito de garantía (margen inicial) en situaciones de apalancamiento. Las estupendas interpretaciones de Jeremy Irons y de Kevin Spacey, sin menosprecio del resto del reparto, hacen que esta película tenga un atractivo especial que nos mantiene en constante atención. Probablemente sea la película en la que mejor se detectan las malas prácticas empleadas en los negocios por los grandes fondos de inversión con ayuda de la ingeniería financiera. Ante un mercado poco regulado, los grandes inversores pueden hundir (o también sobrevalorar de manera artificiosa) cualquier compañía que cotice en bolsa. Uno de los casos más sonados recientemente ha sido del de Gamestop" [Martín-de-Santos, 2021]. Película solemne hasta en la banda musical, sin las acciones rápidas vistas en películas similares. Son necesarios unos conocimientos básicos sobre finanzas para comprender algunas partes de los diálogos que han propiciado críticas negativas. La aparición de elementos femeninos es escasa y se limita a la jefa de la sección de riesgos Sarah Robertson (Demi Moore) con un relativo protagonismo y a la insignificante aparición de Laurent Bratbert (Susan Blackwell), Lucy (Grace Gummer) y diversas ejecutivas de apoyo (Maria Dizzia, …).

22- *Too big to fail* / *Malas noticias* (2011). Película de tendencia documental incluyendo retazos de imágenes obtenidas de noticiarios televisivos. Personajes representando a las personas responsables de las trampas financieras con especial alusión a Henry (Hank en la película) Paulson (William Hurt), secretario del Tesoro durante el gobierno Bush (hijo) por su polémico rescate a la banca. La decisión de la Reserva Federal, ante la falta de acuerdo con el Congreso para la compra de activos bancarios, decidió inyectar dinero a las principales entidades financieras (lo que en definitiva suponía un rescate), incluso a las que no se venían necesitadas de liquidez, con el fin de que estas, a su vez, activaran el crédito a los particulares. Película bien argumentada, pero algo atropellada y vertiginosa que permite poco tiempo para la reflexión. En la Junta de la Reserva Federal tan sólo figura una mujer que apenas interviene en las reuniones. Hay un documental muy similar con entrevistas a Hank Paulson, Timothy Geithner, Ben Bernanke, Larry Summers, George W. Bush, Barack Obama, Janet Yellen, Christine Lagarde, … titulado *Panic: The Untold Story of the 2008 Financial* Crisis (*Pánico, la crisis jamás contada de la crisis del 2008,* 2018).

23- *Le capital* / *El capital* (2012). Aquí se repite el tópico de las opas tal como aparece, por ejemplo, en *Other People's Money*. En el afán de demonizar a la banca, se llega a situaciones poco creíbles. No obstante, los diálogos están muy cuidados y terminan con un final provocativo: "Seguiremos robando a los pobres para dárselo a los ricos". El contenido trata la historia del modesto banquero Marc Tourneuil (Gad Elmaleh) que de modo inopinado acaba por ser presidente de un gran banco de inversión. Pensada su elección como fantoche manejable, pronto adoptará una actitud marcada por su particular personalidad, pero sus objetivos no son menos ambiciosos que los de la anterior presidencia. Es cine de denuncia en el que la mujer apenas tiene presencia más que como esposa (Diane Tourneuil interpretada por Natacha Régnier), hermana (intérprete Aline Stinus) o amante (Nassim / Liya Kebede).

24- *Assault on Wall Street* / *Asalto a Wall Street* (2013). El contenido se centra en la venganza de un guardia de seguridad, Jim Baxford (Dominic Purcell), contra los responsables financieros de la crisis que estalló en 2008 y los abogados oportunistas, y que le produjeron toda clase de desgracias, tanto económicas como sentimentales. Su mujer Roxie Baxford (Erin Karpluk) necesita un caro tratamiento por enfermedad, pero sus ahorros en bonos garantizados se esfuman debido a una malversación por parte del bróker; ella muere y gran parte del personal de la agencia de inversiones es asesinada por Jim, incluido el jefe. A pesar de presentar una fácil venganza poco creíble y de no haber tenido buena aceptación por parte de la crítica, hay que reconocer que involucra al espectador en la acción y desata el instinto de justicia, además de mostrar otros valores como el compañerismo. Se trata de una película minusvalorada en cuanto a la forma, y revolucionaria en cuanto al contenido en el sentido de que recurre a la violencia para compensar el daño sufrido. Tal vez denostada por antisistema y antipatriótica. Las artimañas financieras y las reflexiones de carácter ético están bien descritas y el ritmo apropiado.

25- *Il capitale umano* / *El capital humano* (2013). Contenido abrumador sobre fraudes financieros y asuntos amorosos, tremendamente orientador para las economías domésticas. Magnífica ilación de ida y vuelta entre los cuatro capítulos de la obra. Buena narración de los acontecimientos en torno a un atropello mortal que implica a los miembros de dos familias. Provoca tensión en el espectador la mayor parte del tiempo. Refleja fielmente la realidad financiera internacional en tiempos de crisis mezclada con los trajines familiares y acciones pasionales de amores y desamores. El epílogo hace una peculiar interpretación de lo que es el capital humano en la teoría marxista. Acertado tratamiento de la mujer que prefiere el amor al dinero en una sociedad acosada por fondos buitre y por delirios de grandeza clasista. Reparto de protagonismo entre personajes femeninos y masculinos mitad por mitad, pero bajo el poder económico de los hombres como fuentes de ingresos.

26- *The Wolf of Wall Street* / *El lobo de Wall Street* (2013). Reafirmándome en una valoración anterior (Martín-de-Santos, 2021), es una película fantasiosa, megalómana y sensacionalista. Su larga duración y los diálogos con constantes usos de palabrotas exasperan un poco la sensibilidad del espectador por más que estén pensados para reaccionar contra las artimañas de los agentes de cambio y bolsa vendedores de humo. Ambientación excéntrica como la gestora de inversiones más parecida a una discoteca que a un lugar de trabajo. Los trucos infantiles para la evasión de divisas, la anacrónica música de Simon & Garfunkel en la detención de la plantilla de brokers y otras muchas incongruencias, no explican el éxito de una película que recibió importantes premios e, incluso, varias nominaciones a los óscar. Lujos, drogas y prostitución son recursos que intentan llamar la atención. La mujer es tratada como objeto del deseo y juguete al servicio del dinero.

27- *The big short* / *La gran apuesta* (2015). Excelente película testimonial sobre la actividad económica del médico estadounidense Michael Burry (Christian Bale) y tres colegas que apostaron sus ahorros contra la previsible crisis bancaria causada principalmente por las hipotecas de alto riesgo (*subprime*) y las obligaciones colaterales de deuda colateral o CDOs (*Collateralized Debt Obligation*). Aunque el guion está escrito a partir del libro de Michael Lewis, y a pesar de algunos aspectos caricaturescos sobre Burry, en general la trama se ajusta a la realidad. En pocas palabras: las deudas procedentes de préstamos hipotecarios fueron manipuladas subrepticiamente por la banca transformándolas en bonos garantizados que no ofrecían tal garantía puesto que los prestatarios tenían cada vez más dificultades en devolver los préstamos y las compañías de seguros y reaseguros no contaban con un fondo de reserva suficiente para hacer frente al riesgo. Los personajes secundarios femeninos son Evie (Karen Gillan), Georgia Hale (Melissa Leo), Cynthia Baum (Marisa Tomei), Selena Gómez (Selena Gómez), Margot Robbie (Margot Robbie) y Host (Lara Grice)[3], a pesar de ello los personajes masculinos duplican en número a las actrices.

28- *Equity* / *Equidad* (2016). Esta película marca el principio de una nueva sensibilidad en el campo de las finanzas. Aunque pueda ser tildada de película feminista, va más allá del feminismo; ya el título lo deja entrever. Es curioso que no se mencione la igualdad, sino la equidad, es decir dar a cada uno lo que se merezca. Es película de mujeres, hecha por mujeres, con las mujeres como protagonistas y sin necesidad de anunciar el feminismo a los cuatro vientos o, lo que es lo mismo: siendo película feminista, no se nota que lo es. Este modo de hacer las cosas le da un valor especial y marcará época. Presenta como naturales y ordinarios los fenómenos que pueden afectar de la misma manera a la vida afectiva tanto de ellas como de ellos. Quizá esto pueda parecer algo chocante por la costumbre de ver en el cine a los hombres como bastiones de los negocios. El modo de presentar los hechos se hace con tanta finura que subyuga al espectador, le transforma en copartícipe de las dificultades y da lo mismo que la ambición, la presión o la traición sea masculina o femenina. Los avatares financieros están relatados con mayor claridad que en películas de contenido similar. El final resulta amargo. En alguna ocasión se ha comparado con *Work Girl* / *Armas de mujer* (1988) que, a mi juicio carece de la profundidad sicológica de los personajes en *Equity*. Es recomendable proyectar *Equity* en las aulas con fines educativos, sobre todo para analizar pormenorizadamente los fondos de cobertura, figura financiera de gran riesgo y escasa regulación.

---

[3] No cuento a mujeres de relleno o extras como la bailarina erótica María Frangos o la terapeuta Leslie Castay.

29- *La chute de l'empire américain* / *La caída del imperio americano* (2018). Un doctor en filosofía, Pierre-Paul Daoust (Alexandre Landry) se encuentra por casualidad con dos bolsas de dinero obtenidas de un atraco en el que mueren los asaltantes. Con la ayuda de un experto financiero y de un abogado sin escrúpulos de tipo económico o sexual, deposita el dinero en paraísos fiscales. A su vez, acaba enamorándose de una joven a la que contrata primeramente como prostituta; ambos terminan ayudando a personas marginadas e incluso invitan a la pareja de policías (hombre y mujer) a colaborar en estas tareas. Aunque el calado financiero es evidente, aquí se abandona la anodina tendencia al cansino diálogo en favor de la aventura y la acción. Velada crítica al capitalismo, no especialmente clara. También en el reparto de personajes secundarios, el número de actores duplica al de las actrices.

30- *White Tiger* / *El tigre blanco* (2021). Reflejo del desarrollo capitalista en la India y pintoresca puesta en escena de las costumbres de aquel país. El tigre blanco es símbolo de depredación y de singularidad. El hilo argumental gira en torno al ascenso de Balram Halwai (Adarsh Gourav) que pasa de pobre aldeano a rico empresario. Astuto y ambicioso, entra como chófer al servicio de un empresario y soportando todo tipo de humillaciones, llega a ser el dueño de una flota de coches para el transporte. Se representa una manera de vivir despiadada, una India no exenta de grandes desigualdades y las corrupciones propias de cualquier sistema político. La variada cadena de aventuras hace ameno el seguimiento de un metraje algo elevado. Se plantean diversas dicotomías: casta de las barrigas grandes / casta de las barrigas pequeñas, espacio rural / espacio urbano, pobreza / riqueza, castas bajas /castas altas, mujeres emancipadas / mujeres subyugadas. El trato a la mujer es mayormente vejatorio excepto cuando se trata de mujeres ricas y cultas.

## Resultados

Una sola película se adapta al test de Bechdel (nº 28; 3,3%). Las mujeres protagonistas no se dan más que en dos películas (nº 12 y 28; 6,6%). En las restantes 27 películas las mujeres ejercen papeles secundarios (81%). En 23 películas (69%), los valores positivos de las mujeres prevalecen sobre los negativos. En 6 películas (18%) prevalecen los valores negativos. En la película nº 22 aparece una mujer que no destaca por presentar valores positivos o negativos. En definitiva, la presencia de la mujer en las películas de temas financieros es insignificante.

Esta visión se contradice con la realidad. El contraste es evidente si establecemos una comparación con el hecho de que los puestos más representativos de las finanzas internacionales hayan estado recientemente o estén ocupados por mujeres: Janet Yellen (Secretaria del Tesoro de EEUU) y presidenta de la Reserva Federal); Christine Lagarde (Presidenta del Banco Central Europeo y anteriormente Presidenta del Fondo Monetario Internacional), Kristalina Georgieva (directora gerente del Fondo Monetario Internacional), Elvira Nabiúlina (Presidenta del Banco Central de la Federación Rusa), Carmen Reinhart (economista jefa del Banco Mundial, Jane Fraser (CEO de Citigroup, uno de los grandes bancos de inversión del mundo), ...

En España, según cifras del Instituto Nacional de Estadística (2020), la participación femenina en los puestos de responsabilidad de la Administración Pública y Privada se acerca cada día más al de los hombres. En todos los sectores de la Administración del Estado, más del 40% de Altos Cargos son mujeres. En el sector privado el porcentaje de mujeres disminuye, por ejemplo, en el conjunto de consejos de administración de las empresas del Ibex-35, es menor (27,7%) y en la presidencia y/o consejeras delegadas el número es imperceptible. Pero lo que importa más que estos datos es la evolución que se está produciendo en el relevo de hombres por mujeres: un aumento del 50% en los últimos 7 años en los consejos de administración y una media aproximada de 3 puntos por año en general en todos los grupos. Otros estudios (Grant Thornton, 2021) ofrecen cifras parecidas. En la empresa privada, el porcentaje de mujeres directivas es el 34%, con ligeras variaciones según zonas. Hay que esperar que todas estas cifras se aproximen entre sí más cada día porque, como han demostrado Ostry et al. (2018), la participación laboral de las mujeres en trabajos tradicionalmente desempeñados por los hombres aumenta más de lo que se cree el crecimiento económico y la productividad.

## Conclusiones

A la vista de los resultados obtenidos, se puede constatar que la mayor parte del cine sobre finanzas reviste casi en su totalidad un protagonismo masculino. Las dos películas (nº 12 y 28) en las que la mujer aparece como protagonista, no deja de ser una asalariada más de los dueños del dinero. La relevancia de las actrices en estos dos casos es, por lo tanto, secundaria, sin embargo, la mayoría de ellas desarrolla valores positivos o actitudes ortodoxas.

La aplastante mayoría de protagonistas que intervienen en películas sobre la bolsa son hombres. Este es un buen indicador en favor de las mujeres porque en general lo que exponen estos filmes son denuncias sobre el irregular funcionamiento de determinadas acciones bursátiles guiadas por la ambición y la corrupción.

Últimamente, el protagonismo femenino está adquiriendo gran trascendencia en el cine, y en especial en el de los países menos desarrollados. Si esto tiene algo que ver con la realidad, se debe seguramente a que las mujeres en estos lugares soportan en buena parte el peso de las economías domésticas

En el actual panorama internacional, caracterizado por una economía liberal incontrolada y capaz de provocar los mayores desastres económicos, debidos fundamentalmente al fenómeno de la globalización, no está de más revisar la teoría de la inestabilidad inherente, vinculada a la teoría de los ciclos, para poder prevenir, dentro de lo posible desgracias evitables. En este sentido, se apunta la necesidad de tener más en cuenta la obra de Minsky (1986) y la de dar mayor participación a las mujeres en las decisiones financieras.

## Bibliografía

# ANEJO 1

| TÍTULO | PRO<br>TA<br>GO<br>NIS<br>TA | DE REPARTO | ATRIBUTOS POSITIVOS | ATRIBUTOS NEGATIVOS | TEST DE BECHDEL | OTROS |
|---|---|---|---|---|---|---|
| 1. *Die Finanzen des Großherzogs / Las finanzas del Gran Duque* (1924) | | X | X | | | Mujer coprotagonista la Gran Duquesa Olga |
| 2. *American Madness* (*La locura del dólar,* 1932) | | X | X | | | Alternancia de novias ejemplares: Helen (Constance Cummings) y casadas indecisas: señora Dickson (Kay Johnson) |
| 3. *Modern Times / Tiempos modernos* (1936) | | X | X | | | Joven desahuciada (Paulette Gocdard) |
| 4. *The Grapes of Wrath Las uvas de la ira* (1940) | | X | X | | | Ma Joad (Jane Darwell) madre de Tom (Henry Fonda) obtuvo el Óscar a la mejor actriz secundaria |
| 5. *It's a wonderful life / ¡Qué bello es vivir!* (1946) | | X | X | | | Mary Atch (Donna Reed), personaje secundario, hace pareja con George |
| | | | | | | Bailey (James Stewart). |
| 6. *The Man in the White Suit / El hombre del traje blanco* (1951) | | X | X | | | Daphne (Joan Greenwood) pareja de Sid Stratton (Alec Guinness) |

| TÍTULO | PRO<br>TA<br>GO<br>NIS<br>TA | DE REPARTO | ATRIBUTOS POSITIVOS | ATRIBUTOS NEGATIVOS | TEST DE BECHDEL | OTROS |
|---|---|---|---|---|---|---|
| 7. Bienvenido, Mister Marshall (1953) | | X | X | | | Actuación de Lolita Sevilla impuesta por la productora UNINCI |
| 8. Executive Suite / La torre de los ambiciosos (1954) | | X | X | | | Elenco de actrices de prestigio que actúan como muletas de la acción principal |
| 9. Ziemia obiecana / La tierra de la gran promesa (1975) | | X | X | | | Mujeres, en general, maltratadas en todas las escalas sociales |
| 10. Tradings Places / | | X | | X | | Presencia de Ophelia (Jamie Lee Curtis) en |
| Entre pillos anda el juego (1983) | | | | | | calidad de prostituta |
| 11. Wall Street (1987) | | X | | X | | Darien Taylor (Dary Hannah) mujer calculadora entre dos hombres |
| 12. Working Girl / Armas de mujer (1988) | X | | X | | | Tess McGill (Melanie Grfiith) es coprotagonista |

| TÍTULO | PRO TA GO NIS TA | DE REPARTO | ATRIBUTOS POSITIVOS | ATRIBUTOS NEGATIVOS | TEST DE BECHDELL | OTROS |
|---|---|---|---|---|---|---|
| 13. Other People's Money / Con el dinero de los demás (1991) | | X | X | | | Interviene también Anna Nicole Smith, chica Play Boy |
| 14. The Hudsucker Proxy / El gran salto (1994) | | X | X | | | Amy Archer (Jennifer Jason Leigh), periodista |
| 15. Rogue Trader / El gran farol (1999) | | X | x | | | Lisa Leeson (Anna Friel), esposa del protagonista |
| 16. Boiler Room / El informador | | X | X | | | secretaria Abbie (Nia Long**)** |
| (2000) | | | | | | |
| 17. A Beautiful Mind / Una mente maravillosa (2001) | | X | X | | | Alicia Nash (Jennifer Connelly) alumna casada con John Forbes Nash (Russell Crowe) |
| 18. Die Fälscher / Los falsificadores (2007) | | | | | | Ausencia de personajes femeninos |

—

| TÍTULO | PRO<br>TA<br>GO<br>NIS<br>TA | DE REPARTO | ATRIBUTOS POSITIVOS | ATRIBUTOS NEGATIVOS | TEST DE BECHDELL | OTROS |
|---|---|---|---|---|---|---|
| 19. The Company men (2010) | | X | X | X | | Maggie (Rosemarie DeWit) incorporación laboral.<br>Sally Wilcox (María Bello), amante |
| 20. The Last Days of Lehman Brothers / Los últimos días de Lehman Brothers (2010) | | X | X | | | Ezzy (Katrene Rochell) y Donna Lewis (Laura Brook) apenas actúan |
| 21, Margin Call / El precio de la codicia | | X | X | | | Sarah Robertson (Demi Moore); |
| (2011) | | | | | | Laurent Bratbert (Susan Blackwell) escasa participación |
| 22. Too Big to Fall / Malas noticias (2011) | | X | | | | Michele Davis (Cynthia Nixon)<br>*Wendy Paulson (Kathy Baker)* imperceptible participación |
| 23. Le capital / El capital (2012) | | X | | X | | La mujer más destacada es Nassim (Liya Kebede) amante del protagonista |
| 24. Assault on Wall Street / Asalto a Wall Street (2013) | | X | X | | | Roxie Baxford (Erin Karpluk) esposa del protagonista |

| TÍTULO | PRO<br>TA<br>GO<br>NIS<br>TA | DE REPARTO | ATRIBUTOS POSITIVOS | ATRIBUTOS NEGATIVOS | TEST DE BECHDELL | OTROS |
|---|---|---|---|---|---|---|
| 25. Il capitale umano / El capital humano (2013) | | X | X | | | Excelente actuación de Carla Bernaschi (Valeria Bruni Tedeschi) |
| 26. The Wolf of Wall Street / El | | X | | X | | Naomi Lapaglia (Margot Robbie) esposa de Jordan |
| lobo de Wall Street (2013) | | | | | | Belford (Leonardo DiCaprio) |
| 27. The Big Short / La gran apuesta (2015) | | X | X | | | Margot Robbie expone clara y sucintamente el fraude de los bonos basura |
| 28. Equity / Equidad | X | | X | | X | Excelente interpretación de Naomi Bishop (Anna Gunn) |
| 29. La chute de l'empire américain / La caída del imperio americano (2018) | | X | | X | | Destacan Aspasie/Camille Lafontaine (Maripier Morin) prostituta |
| 30. White Tiger / El tigre blanco (2021) | | X | X | | | Pinky Madam (Priyanka Chopra) mujer liberal |